# Transport equation for the drift velocity in predicting the filtered Eulerian drag force: a theoretical development

**Xiao Chen and Ming Jiang**

School of Chemical Engineering and Technology, Xi'an Jiaotong University, Xi'an 710049, China

**Qiang Zhou**

State Key Laboratory of Multiphase Flow in Power Engineering, Xi'an Jiaotong University, Xi'an 710049, China

School of Chemical Engineering and Technology, Xi'an Jiaotong University, Xi'an 710049, China

**Significance**

The drift velocity has been proven to have significant relevance to the filtered Eulerian drag force by numerous correlative analyses of fully resolved simulations. It is a sub-grid quantity defined as the difference between the filtered gas velocity seen by the particle phase and the resolved filtered gas velocity. In literature, it is shown that various algebraic models for the drift velocity fail to give entirely satisfactory prediction of the filtered drag force. Unlike previous works, we theoretically derived the transport equation for the drift velocity from the standard two-fluid model (TFM) equations without any additional assumptions. The new approach, though requires additional closures for the new unresolved terms, is believed to have the potential to improve the precision of the estimation of this sub-grid quantity and hence give better prediction of the filtered drag force for gas-solid fluidization problems.

**Key words**: Fluidization, Multi-phase flow, Two-fluid model, Coarse-grid simulations, Drift velocity

## Introduction

Industrial-scale fluidized beds contain trillions of particles with complicated gas-particle flow structures. These flows exhibit a wide range of spatial scales: from a few particle diameters to the scale of vessels. The multi-scale nature of fluidization demands the microscopic details to accurately predict the hydrodynamics of gas-solid flows, which is often unaffordably expensive to resolve even in the efficient two-fluid model (TFM, also known as Euler-Euler model) simulations[1,2]. Thus, to overcome this problem, one of the most promising approaches, called filtered two-fluid model (fTFM), is being developed to simulate large-scale flow problems[3-8]. The idea of fTFM is similar to large eddy simulation (LES) of turbulent single-phase flows. In fTFM, only the filtered transport equations on the "coarse-grid" (which is not fine enough to resolve the particle clustering and other micro quantities at small scale) are solved and the effects of unresolved structures at the sub-grid are modeled.

Previous research work has implied that during all the constitutive terms of filtered transport equations, the filtered drag force is the most important one[4,5,8]. Many researchers have attempted to model the sub-grid contribution of the drag term: Igci & Sundaresan[7] proposed a filtered drag force as a function of the filtered volume fraction and the filter size. Filtered slip velocity was later added in as an additional factor[8,9]. The drift velocity was first proposed in predicting filtered drag force by Parmentier et al.[8]. And then Ozel et al.[4] demonstrated that the drift velocity was of significant importance in accurately estimating solid flux in a 3D periodic channel flow. Later, Ozel et al.[10] showed distinct influence of the drift velocity on filtered drag force. And these works are consistent with lattice-Boltzmann-discrete-element-method study[11] with much smaller systems and TFM simulations with much larger systems[4,8].

Though it has been widely demonstrated that the drag correction is well correlated by the drift velocity in many gas-solid flows[12], the drift velocity is not readily available in coarse-grid simulations and thus needs additional designed closures. Among the closures proposed for the drift velocities, the scale similarity model, which is viewed as the best available approach, is still shown to be not entirely satisfactory in predicting the filtered drag force[10,12]. It is believed that solving an additional transport equation for the drift velocity has the potential to improve the precision of the estimation of this sub-grid quantity. Indeed, the use of a transport equation for better prediction of the drift velocity has been suggested in the recent literature[10-12]. However, the formulation of the transport equation is still yet to be done, probably due to the complex mathematical expression of the drift velocity in its original definition. This motivates us and the main task of the present work is to give the rigorous theoretical development of the desirable transport equation. It is mathematically shown that, the product of drift velocity and volume average of solid volume fraction can be represented by the volume average of the Favre fluctuation of the gas phase velocity. Thus the transport equation for the drift velocity could be easily obtained by deriving the governing equation for the volume average of the Favre fluctuation of the gas phase velocity.

Without the use of any additional assumptions, the transport equation for the drift velocity from the standard Euler-Euler transport equations is derived. The physical meaning of the new terms appearing in the equation are also analyzed and interpreted. This letter aims to solely lay the foundation for the new approach with the assistance of a transport equation to obtaining the most prominent sub-grid quantity, i.e., the drift velocity. Closures for the new unresolved terms are left for future work.

The paper is organized as follows: the next section gives the introduction of filtering procedure and filtered TFM equations; after that the transport equation for the drift velocity is derived and analyzed in detail; and the principal results are summarized at the end.

**Filtering procedure and filtered equations**

In TFM, the continuity and momentum balance equations for the gas phase are as follows[3,13]:

$$\frac{\partial}{\partial t}(\rho_g \phi_g) + \frac{\partial}{\partial x_j}(\rho_g \phi_g u_{g,j}) = 0 \qquad (1)$$

$$\frac{\partial}{\partial t}(\rho_g \phi_g u_{g,i}) + \frac{\partial}{\partial x_j}(\rho_g \phi_g u_{g,i} u_{g,j}) = -\phi_g \frac{\partial \sigma_{g,ij}}{\partial x_j} - f_{gs,i} + \rho_g \phi_g g_i \qquad (2)$$

In Eqs. (1) and (2), $\rho_g$, $\phi_g$, $u_{g,i}$ and $\sigma_{g,ij}$ are the density, volume fraction, velocity and stress tensor of the gas phase, respectively, where the subscript *i* and *j* are the direction indices of the Cartesian coordinate system. $f_{gs,i}$ is the interaction force between the gas and solid phases per unit volume.

To obtain the filtered model equations, the simple volume-averaging with a box filter[9,11] is employed, and the filtered part of any variable $\psi$ could be written as

$$\bar{\psi} = \iiint_{\substack{filtering \\ window}} \psi G(x, y, z) dx dy dz \qquad (3)$$

where the weighting function $G(x, y, z)$ satisfies $\iiint_{\substack{filtering \\ window}} G(x, y, z) dx dy dz = 1$. With this operation, the fluctuating component of volume-averaged variable $\bar{\psi}$ could be obtained as follows

$$\psi' = \psi - \bar{\psi} \qquad (4)$$

Further using the same filter, the Favre-averaged flow variables (with a tilde) and the corresponding Favre fluctuations (with double prime) could be defined as:

$$\tilde{\psi}_k = \frac{\overline{\phi_k \psi_k}}{\overline{\phi_k}} \tag{5}$$

$$\psi_k'' = \psi_k - \tilde{\psi}_k \tag{6}$$

where the subscript $k=s, g$ denotes the solid or gas phase separately. For the above averaging, the following identities apply

$$\overline{\psi'} = 0 \tag{7}$$

$$\overline{\phi_k \mathbf{u}_k''}\big|_{k=g,s} = 0 \tag{8}$$

$$\overline{\psi}_k + \psi_k' = \tilde{\psi}_k + \psi_k'' \tag{9}$$

Using the filtering technique, the derivation of the filtered continuity and momentum balance equations for gas phase could be found in the literature[3,5], we simply summarize the results as:

$$\frac{\partial}{\partial t}(\rho_g \overline{\phi}_g) + \frac{\partial}{\partial x_j}(\rho_g \overline{\phi}_g \tilde{u}_{g,j}) = 0 \tag{10}$$

$$\frac{\partial}{\partial t}(\rho_g \overline{\phi}_g \tilde{u}_{g,j}) + \frac{\partial}{\partial x_j}(\rho_g \overline{\phi}_g \tilde{u}_{g,j} \tilde{u}_{g,i}) = -\overline{\phi}_g \frac{\partial \overline{\sigma}_{g,ij}}{\partial x_j} - \rho_g \frac{\partial}{\partial x_j}(\overline{\phi_g u_{g,j}'' u_{g,i}''}) \\ - (\overline{f}_{gs,i} + \overline{\phi_g' \frac{\partial \sigma_{g,ij}'}{\partial x_j}}) + \rho_g \overline{\phi}_g g_i \tag{11}$$

Following the studies by Parmentier et al.[8] and Ozel et al.[10], the sub-grid drift velocity is defined as

$$\overline{\phi}_s \tilde{v}_{d,i} = \overline{\phi_s(u_{g,i} - u_{s,i})} - \overline{\phi}_s(\tilde{u}_{g,i} - \tilde{u}_{s,i}) = \overline{\phi_s u_{g,i}} - \overline{\phi}_s \tilde{u}_{g,i} \tag{12}$$

where $\phi_s$ is the solid volume fraction and $u_{s,i}$ is the solid velocity at direction $i$. Considering $\phi_s = 1 - \phi_g$ and using the definition (5), we have

$$\overline{\phi_s \tilde{v}_{d,i}} = \overline{u}_{g,i} - \overline{\phi_s u_{g,i}} - \overline{\phi_s \tilde{u}_{g,i}} = \overline{u}_{g,i} - \overline{\phi}_g \tilde{u}_{g,i} - \overline{\phi}_s \tilde{u}_{g,i} = \overline{u}_{g,i} - \tilde{u}_{g,i} \quad (13)$$

Take the volume-average of Eq. (9), and replace $\psi_k$ with $u_{g,i}$ in the obtained equation yielding

$$\overline{u}_{g,i} - \tilde{u}_{g,i} = \overline{u''_{g,i}} \quad (14)$$

Again, using Eq. (9) and above equation (14) could give us

$$\overline{u''_{g,i}} = u''_{g,i} - u'_{g,i} \quad (15)$$

and hence the definition of the drift velocity can be simply represented as

$$\overline{\phi_s \tilde{v}_{d,i}} = \overline{u''_{g,i}} \quad (16)$$

Therefore, the problem of finding the transport equation for $\tilde{v}_{d,i}$ has been transformed into the problem of finding the transport equation for $\overline{u''_{g,i}}$, which is the volume average of the Favre fluctuation of the gas phase velocity.

**Transport equation for the drift velocity**

It has been shown that $\tilde{v}_{d,i}$ is closely related to $\overline{u''_{g,i}}$ via Eq. (16). Thus, the transport equation for $\overline{u''_{g,i}}$ would be equivalent to the wanted transport equation for $\tilde{v}_{d,i}$. In what follows, we show that the transport equation for $\overline{u''_{g,i}}$ could be obtained from the difference between the transport equation for $\tilde{u}_{g,i}$ and the transport equation for $\overline{u}_{g,i}$ as indicated by Eq. (13).

Multiply Eq. (10) by $u_{g,i}$, then subtract the resulted equation from Eq. (11) yielding

$$\rho_g \overline{\phi}_g \frac{\partial}{\partial t}(\tilde{u}_{g,i}) + \rho_g \overline{\phi}_g \tilde{u}_{g,j} \frac{\partial}{\partial x_j}(\tilde{u}_{g,i}) = -\overline{\phi}_g \frac{\partial \overline{\sigma}_{g,ij}}{\partial x_j} - \rho_g \frac{\partial}{\partial x_j}(\overline{\phi_g u''_{g,j} u''_{g,i}})$$
$$- (\overline{f}_{gs,i} + \overline{\phi'_g \frac{\partial \sigma'_{g,ij}}{\partial x_j}}) + \rho_g \overline{\phi}_g g_i \quad (17)$$

Similarly, subtract Eq. (1) multiplied by $u_{g,i}$ from Eq. (2), then divide the result by $\phi_g$ giving

$$\rho_g \frac{\partial}{\partial t}(u_{g,i}) + \rho_g u_{g,j} \frac{\partial}{\partial x_j}(u_{g,i}) = -\frac{\partial \sigma_{g,ij}}{\partial x_j} - \frac{f_{gs,i}}{\phi_g} + \rho_g g_i \tag{18}$$

Take the volume average of Eq. (18), we have

$$\rho_g \frac{\partial}{\partial t}(\bar{u}_{g,i}) + \rho_g \bar{u}_{g,j} \frac{\partial}{\partial x_j}(\bar{u}_{g,i}) = -\frac{\partial \bar{\sigma}_{g,ij}}{\partial x_j} - \rho_g \overline{u'_{g,j} \frac{\partial}{\partial x_j}(u'_{g,i})} \\ - \frac{\overline{f_{gs,i}}}{\bar{\phi}_g} + \frac{1}{\bar{\phi}_g} \overline{\frac{f_{gs,i}}{\phi_g} \phi'_g} + \rho_g g_i \tag{19}$$

Replacing $\bar{u}_g$ by $\tilde{u}_g + \overline{u''_g}$ (from Eq. (14)) in Eq. (19), and multiplying the obtained equation by $\bar{\phi}_g$ gives

$$\rho_g \bar{\phi}_g \frac{\partial}{\partial t}(\tilde{u}_{g,i}) + \rho_g \bar{\phi}_g \tilde{u}_{g,j} \frac{\partial}{\partial x_j}(\tilde{u}_{g,i}) + \rho_g \bar{\phi}_g \frac{\partial}{\partial t}(\overline{u''_{g,i}}) + \rho_g \bar{\phi}_g \tilde{u}_{g,j} \frac{\partial}{\partial x_j}(\overline{u''_{g,i}}) \\ + \rho_g \bar{\phi}_g \overline{u''_{g,j}} \frac{\partial}{\partial x_j}(\tilde{u}_{g,i}) + \rho_g \bar{\phi}_g \overline{u''_{g,j}} \frac{\partial}{\partial x_j}(\overline{u''_{g,i}}) \\ = -\bar{\phi}_g \frac{\partial \bar{\sigma}_{g,ij}}{\partial x_j} - \rho_g \bar{\phi}_g \overline{u'_{g,j} \frac{\partial}{\partial x_j}(u'_{g,i})} - \overline{f_{gs,i}} + \overline{\frac{f_{gs,i}}{\phi_g} \phi'_g} + \rho_g \bar{\phi}_g g_i \tag{20}$$

Then, subtracting Eq. (17) from Eq. (20), replacing $u'_g$ by $u''_g - \overline{u''_g}$, and considering that $\overline{\overline{u''_g}} = \overline{u''_g}$, we arrive at the transport equation of $\overline{u''_{g,i}}$ in the non-conservative form as follows

$$\rho_g \bar{\phi}_g \frac{\partial}{\partial t}(\overline{u''_{g,i}}) + \rho_g \bar{\phi}_g \tilde{u}_{g,j} \frac{\partial}{\partial x_j}(\overline{u''_{g,i}}) + \rho_g \bar{\phi}_g \overline{u''_{g,j}} \frac{\partial}{\partial x_j}(\tilde{u}_{g,i}) \\ = \rho_g \frac{\partial}{\partial x_j}(\overline{\phi_g u''_{g,j} u''_{g,i}}) - \rho_g \bar{\phi}_g \overline{u''_{g,j} \frac{\partial}{\partial x_j}(u''_{g,i})} + \overline{\frac{f_{gs,i}}{\phi_g} \phi'_g} + \overline{\phi'_g \frac{\partial \sigma'_{g,ij}}{\partial x_j}} \tag{21}$$

It should be mentioned that the box filter adopted in this work gives $\overline{\overline{u''_g}} = \overline{u''_g}$ and this might not be valid under other types of filter, such as a Gaussian filter, in which the assumption from Jiménez et al.[14] could be used to obtain Eq. (21). The first two terms in the right-hand-side in Eq. (21) could be rearranged as follows

$$\rho_g \frac{\partial}{\partial x_j}(\overline{\phi_g u''_{g,j} u''_{g,i}}) - \rho_g \overline{\phi_g u''_{g,j} \frac{\partial}{\partial x_j}(u''_{g,i})}$$

$$= \rho_g \frac{\partial}{\partial x_j}(\bar{\phi}_g \overline{u''_{g,j} u''_{g,i}}) + \rho_g \frac{\partial}{\partial x_j}(\overline{\phi'_g u''_{g,j} u''_{g,i}})$$

$$- \left[ \rho_g \bar{\phi}_g \frac{\partial}{\partial x_j}(\overline{u''_{g,j} u''_{g,i}}) - \rho_g \bar{\phi}_g \overline{u''_{g,i} \frac{\partial}{\partial x_j}(u''_{g,j})} \right] \quad (22)$$

$$= \rho_g \frac{\partial}{\partial x_j}(\overline{\phi'_g u''_{g,j} u''_{g,i}}) + \rho_g \overline{u''_{g,j} u''_{g,i}} \frac{\partial}{\partial x_j}(\bar{\phi}_g) + \rho_g \bar{\phi}_g \overline{u''_{g,i} \frac{\partial}{\partial x_j}(u''_{g,j})}$$

Substitute Eq. (22) in Eq. (21) and rearrange Eq. (21) yielding

$$\rho_g \bar{\phi}_g \frac{\partial}{\partial t}(\overline{u''_{g,i}}) + \rho_g \bar{\phi}_g \tilde{u}_{g,j} \frac{\partial}{\partial x_j}(\overline{u''_{g,i}}) = -\rho_g \bar{\phi}_g \overline{u''_{g,j}} \frac{\partial}{\partial x_j}(\tilde{u}_{g,i}) + \overline{\frac{f_{gs,i}}{\phi_g} \phi'_g}$$

$$+ \rho_g \frac{\partial}{\partial x_j}(\overline{\phi'_g u''_{g,j} u''_{g,i}}) + \rho_g \overline{u''_{g,j} u''_{g,i}} \frac{\partial}{\partial x_j}(\bar{\phi}_g) + \rho_g \bar{\phi}_g \overline{u''_{g,i} \frac{\partial}{\partial x_j}(u''_{g,j})} \quad (23)$$

$$+ \frac{\partial(\overline{\sigma'_{g,ij} \phi'_g})}{\partial x_j} - \overline{\sigma'_{g,ij} \frac{\partial \phi'_g}{\partial x_j}}$$

Add Eq. (10) multiplied by $\overline{u''_{g,i}}$ to Eq. (23) and rearrange the resulted equation giving the transport equation for $\overline{u''_{g,i}}$ in the conservation form as follows

$$\frac{\partial}{\partial t}(\rho_g \bar{\phi}_g \overline{u''_{g,i}}) + \frac{\partial}{\partial x_j}(\rho_g \bar{\phi}_g \tilde{u}_{g,j} \overline{u''_{g,i}}) =$$

$$\underbrace{-\rho_g \bar{\phi}_g \overline{u''_{g,j}} \frac{\partial(\tilde{u}_{g,i})}{\partial x_j}}_{(a_1)} + \underbrace{\rho_g \overline{u''_{g,j} u''_{g,i}} \frac{\partial(\bar{\phi}_g)}{\partial x_j}}_{(a_2)} - \underbrace{\overline{\sigma'_{g,ij} \frac{\partial \phi'_g}{\partial x_j}}}_{(a_3)} + \underbrace{\overline{f_{gs,i} \frac{\phi'_g}{\phi_g}}}_{(b)} + \underbrace{\rho_g \bar{\phi}_g \overline{u''_{g,i} \frac{\partial}{\partial x_j}(u''_{g,j})}}_{(c)} \quad (24)$$

$$+ \underbrace{\rho_g \frac{\partial}{\partial x_j}(\overline{\phi'_g u''_{g,j} u''_{g,i}})}_{(d_1)} + \underbrace{\frac{\partial(\overline{\sigma'_{g,ij} \phi'_g})}{\partial x_j}}_{(d_2)}$$

Since $\bar{\phi}_s \tilde{v}_{d,i} = \overline{u''_{g,i}}$, Eq. (24) is indeed the wanted transport equation for the drift velocity. This equation does not contain the gravity term, indicating that the gravity acceleration has no effect on the drift velocity and hence will not affect the filtered drag force. This is consistent with the recent finding obtained by Ozel et al.[10] via numerical simulations. The physical meaning of the various right-hand-side terms in Eq. (24) is as follows:

($a_1$) production term due to large-scale shear.

($a_2$) production term due to large-scale gradient of the gas volume fraction.

($a_3$) production term due to micro-scale gradient of the gas volume fraction fluctuations. This term appears as the correlation between gas-phase stress fluctuations and the gas volume fraction fluctuations. This term can be further split into two parts since $\sigma'_{g,ij} = -p'_g \delta_{ij} + \tau'_{g,ij}$.

($b$) source term accounting for the contribution of the inhomogeneous drag force at the microscopic scale. This term appears as the correlation between the microscopic drag force and the relative gas volume fraction fluctuations.

($c$) correlation between velocity fluctuation and dilatation fluctuation. This term is specific to non solenoidal fluctuation velocity fields.

($d_1$) turbulent diffusion due to the turbulent flux $\overline{\phi'_g u''_{g,j} u''_{g,i}}$.

($d_2$) diffusion due to gas-phase stress fluctuation. This term can be further split into pressure diffusion and viscous diffusion since the identity $\overline{\sigma'_{g,ij} \phi'_g} = -\overline{p'_g \phi'_g} \delta_{ij} + \overline{\tau'_{g,ij} \phi'_g}$ applies.

An important aspect of the transport equation is that the production term ($a_1$) is represented exactly, so it does not require modeling. It is also noted that, the other six terms in the right-hand-side in Eq. (24) do need additional closures for estimating the drift velocity. To formulate closures for terms except ($a_3$), strategies adopted for compressible single-phase turbulent flow (e. g. Taulbee & Vanosdol[15]) could be followed. However, term ($a_3$) does not have a counterpart in single-phase governing equations since its presence is due to the inter-phase interaction between the gas and solid phases. To simplify the modeling for this term, the simple assumption of $\|\phi'_g / \overline{\phi}_g\| \ll 1$ could be applied as follows

$$\overline{f_{gs,i} \frac{\phi'_g}{\phi_g}} = \frac{\overline{f_{gs,i}\phi'_g}}{\overline{\phi}_g + \phi'_g} \approx \frac{\overline{f_{gs,i}\phi'_g}}{\overline{\phi}_g} = \frac{\overline{f'_{gs,i}\phi'_g}}{\overline{\phi}_g} \tag{25}$$

Then the flux $\overline{f'_{gs,i}\phi'_g}$ could be modeled by the conventional gradient model. It should be noted that this simplification is not always valid especially at the borders of bubbles or clusters. Thus a more sophisticated modeling strategy of $\overline{(f_{gs,i}\phi'_g)/\phi_g}$ may be required in large scale simulations with significant inhomogeneity. We end our discussion on the modeling here since the formulation and the validation of the closures for all the unresolved terms in Eq. (25) needs continuous and synergistic efforts from the research community in this field.

**Summary**

The transport equation of the drift velocity, which is believed to help estimate the filtered fluid-particle interaction force with high precision in fluidization problems, has been developed in this letter without additional assumption from standard TFM transport equations (1) and (2). Since the product of drift velocity and volume average of solid volume fraction ($\overline{\phi}_s \tilde{v}_{d,i}$) is mathematically equivalent to the volume average of the Favre fluctuation of the gas phase velocity ($\overline{u''_{g,i}}$), we obtain the wanted equation via formulating a transport equation for $\overline{u''_{g,i}}$.

The production term due to large-scale shear in the wanted transport equation (24) can be exactly obtained without modeling in coarse grid simulations, while other unsolved terms need additional closures. This clearly requires future efforts from the community in this field. The importance of the transport equation for the drift velocity in fTFM is believed to resemble that of the transport equations for Reynolds stresses in single-phase Reynods-averaged Navier Stokes (RANS) turbulence simulations, and it has been demonstrated that the transport equations for Reynolds stress gives better prediction of Reynolds stresses than conventional algebraic Reynolds-stress models. Thus, fTFM assisted by the transport equation for the drift velocity would be a promising way in studying large-scale gas-solid fluidization problems.

**Acknowledgements**

We are grateful to the financial support by the National Natural Science Foundation of China (91634114) and the Natural Science Foundation of Jiangsu Province for Youth (BK20160390).

AIAA-91-0524.